\documentclass{elsart}
\usepackage{natbib}

\def\apj{ApJ}

\def\apjs{ApJS}
\def\apss{Ap\&SS}
\def\aap{A\&A}

\def\mnras{MNRAS}
\def\lesssim{\mathrel{\hbox{\rlap{\hbox{\lower4pt\hbox{$\sim$}}}\hbox{$<$}}}}
\def\gtrsim {\mathrel{\hbox{\rlap{\hbox{\lower4pt\hbox{$\sim$}}}\hbox{$>$}}}}

\def\astrobj1{G117-B15A}
\def\astrobj2{L19-2}
\def\astrobj3{R~548}
\def\astrobj4{G29-38}

\begin{document}

\runauthor{C\'orsico et al.}

\begin{frontmatter}
\title{The potential  of the  variable DA white  dwarf G117-B15A  as a
tool for Fundamental Physics}

\author[laplata]{Alejandro H. C\'orsico\thanksref{1},}
\author[laplata]{Omar G. Benvenuto\thanksref{2},}
\author[laplata]{Leandro G. Althaus\thanksref{3},}
\author[barcelona1]{Jordi Isern\thanksref{4},}
\author[barcelona2]{and Enrique Garc\'\i a--Berro\thanksref{4}}
                   
\address[laplata]{Facultad de Ciencias Astron\'omicas y
                   Geof\'{\i}sicas,\\  
                   Universidad Nacional de La Plata,\\ 
                   Paseo del Bosque S/N,\\ (1900) 
                   La Plata, Argentina\\
                   e-mails: acorsico, obenvenuto, 
                   althaus@fcaglp.fcaglp.unlp.edu.ar\\}
                    
\address[barcelona1]{Institut de Ci\`encies de l'Espai, C.S.I.C.,\\
                    Edifici  Nexus, Gran Capit\`a 2-4,\\ 
                    08034 Barcelona, Spain\\
                    e-mail: isern@ieec.fcr.es\\}

\address[barcelona2]{Departament de F\'\i sica Aplicada,\\  
                    Universitat Polit\`ecnica de Catalunya,\\  
                    Jordi Girona Salgado S/N, M\`odul B-4, Campus Nord,\\
                    08034 Barcelona, Spain\\
                    e-mail: garcia@fa.upc.es}

\thanks[1]{Fellow of the Consejo Nacional de Investigaciones 
           Cient\'{\i}ficas y T\'ecnicas (CO\-NICET)}
\thanks[2]{Member of the Carrera del Investigador
           Cient\'{\i}fico, Comisi\'on de Investigaciones 
           Cient\'{\i}ficas de la Provincia de Buenos Aires (CIC)}
\thanks[3]{Member of the Carrera del Investigador
           Cient\'{\i}fico, Consejo Nacional de Investigaciones 
           Cient\'{\i}ficas y T\'ecnicas (CO\-NICET)}
\thanks[4]{Institut d'Estudis Espacials de Catalunya}

\begin{abstract}
White dwarfs are well studied objects.  The relative simplicity of their
physics  allows to obtain very detailed  models which can be  ultimately
compared with their observed  properties.  Among white dwarfs there is a
specific  class  of  stars,  known  as  ZZ-Ceti  objects,  which  have a
hydrogen-rich  envelope  and show  periodic  variations  in their  light
curves.  G117-B15A   belongs  to  this  particular  set  of  stars.  The
luminosity  variations  have  been  successfully  explained  as  due  to
$g$-mode pulsations.  G117-B15A has been recently claimed to be the most
stable optical clock ever found, being the rate of change of its 215.2 s
period very small:  $\dot{P}=  (2.3 \pm 1.4)  \times  10^{-15}  {\rm s\,
s}^{-1}$,  with a  stability  comparable  to  that  of the  most  stable
millisecond  pulsars.  The  rate of  change  of the  period  is  closely
related to its cooling timescale, which can be accurately  computed.  In
this paper we study the  pulsational  properties of G117-B15A and we use
the observed rate of change of the period to impose  constraints  on the
axion  emissivity and, thus, to obtain a preliminary  upper bound to the
mass of the axion.  This upper bound turns out to be $4\cos^{2}{\beta}\;
{\rm  meV}$ at the 95\%  confidence  level.  Although  there  are  still
several  observational and theoretical  uncertainties,  we conclude that
G117-B15A is a very promising  stellar  object to set up  constraints on
particle physics.
\end{abstract}

\begin{keyword}
Elementary particles: astronomical observations;
Elementary particles: astrophysics;
stars: oscillations;
stars: white dwarfs.
\PACS 95.85.R \sep 95.30.C \sep 97.10.S \sep 97.20.R
\end{keyword}

\end{frontmatter}


\vskip 1cm

\section{Introduction} \label{sec_introd}

As  already  pointed  out in the  excellent  review of  Raffelt  (2000),
astrophysical and cosmological  arguments and observations have become a
well  known  tool to obtain  empirical  information  or  constraints  on
existing  or  hypothetical   elementary   particles.  One  of  the  most
important  reasons  for this is that the dense  environment  of stars is
potentially a powerful source of low-mass weakly interacting  particles.
Since these particles  subsequently  escape from the star this mechanism
constitutes  a sink of  energy  that  ultimately  modifies  the  stellar
lifetimes, thus allowing a comparison with the observed lifetimes.  This
is  particularly  useful  since,  as it is  well  known,  the  different
non-standard  theories  leave open the  possibility  that several exotic
particles (like axions or gravitons) could exist.  Moreover, for several
of these  particles  there  are not yet  laboratory  experiments  in the
relevant mass range that could  eventually  impose tight  constraints on
their existence.

Among other weakly  interacting  massive  particles, axions are the most
promising  candidates  for  non-baryonic  dark matter and,  therefore, a
great deal of attention  has been paid to these  particles.  However the
window where axions are viable dark matter  candidates is  progressively
getting  smaller  (Bergstr\"om  2000).  Axions were  proposed  more than
twenty years ago as a solution to the strong CP problem (Peccei \& Quinn
1977), and have a small mass $m_{\rm ax}=0.60\, {\rm eV} \, 10^7\, ({\rm
GeV}/f_{\rm  a})$, where $f_{\rm a}$ is the  Peccei-Quinn  scale.  Thus,
the  phenome\-no\-logy  of  axions  is  determined  by only  one  number
$(f_{\rm  a})$, the scale of symmetry  breaking.  There are two types of
axion  models  the KVSZ  model  (Kim  1979)  and the DFSZ  model  (Dine,
Fischler \& Srednicki 1981).  In the KVSZ model axions couple to hadrons
and  photons  wereas in the DFSZ  model  axions  also  couple to charged
leptons.  The coupling  strengths depend on the specific  implementation
of the Peccei-Quinn mechanism through dimensionless  coupling constants.
Both models do not set any  constraints on the value of $f_{\rm a}$ and,
therefore,  the limits on the mass of the axion  must be  obtained  from
experimental   tests.  The  mass  of  axion  has  been   constrained  by
laboratory  searches but, up to date, there are few  constraints  coming
from accelerator tests in the relevant mass range (Raffelt 2000).

Given this  situation,  stars and, more  specifically,  our Sun has been
widely  used to set up  constraints  on the mass of the  axion.  Several
experiments to detect solar and galactic  axions have been done (Sikivie
1983,  Lazarus et al.  1992) or are  currently  under way (van Bibber et
al.  1994, Matsuki et al.  1996, Hagmann et al.  1998).  Until now these
experiments  have  failed to  detect  axions  and,  consequently,  their
existence is still an  attractive  but somehow  speculative  hypothesis.
The most tight  constraints to the mass of the axion available  nowadays
come  mostly  from  astrophysical  arguments  and allow for a mass range
between  $10^{-2}$  and  $10^{-5}$  eV.  To  be  precise,  most  of  the
astrophysical  limits are on the  interaction  strength of axions either
with   photons   ---   through   the   Primakov   conversion    ($\gamma
\leftrightarrow  a$) in the electric  field of electrons  and nuclei ---
with electrons --- through the Compton ($\gamma + {\rm e}^-  \rightarrow
{\rm e}^- + a$) and Bremsstrahlung  (${\rm e}^- + (A,Z) \rightarrow {\rm
e}^- + (A,Z) + a$)  processes  --- or with  nucleons.  Examples of these
astrophysical arguments on the coupling strengths are the following (see
Raffelt 2000, for an extensive review).  The coupling strength of axions
to photons has been restricted through helioseismological constraints on
solar energy losses  (Degl'Innocenti  et al.  1997), the red  supergiant
Betelgeuse  (Carlsson  1995),  or  the  dynamics  of  the  explosion  of
supernova SN1987a (Burrows,  Ressell \& Turner 1990).  In the DFSZ model
this  leads to an  upper  limit  on the  mass of the  axion  of  $m_{\rm
ax}\lesssim  0.4$ eV.  On its hand the  coupling  strength  of axions to
electrons  within  the  DFSZ  model  is  defined  using a  dimensionless
coupling    constant    $g_{\rm    ae}=2.83\times     10^{-11}    m_{\rm
ax}/\cos^2\beta$, where $\cos\beta$ is a model-dependent  parameter that
it is usually set equal to unity.  The most  restrictive  bounds on this
coupling strength come from the delay of helium ignition in low-mass red
giants  (Raffelt  \& Weiss  1995)  and give an upper  limit  of  $m_{\rm
ax}\cos^2\beta  \lesssim 0.01$ eV.  Finally, the axion-nucleon  coupling
strength has been mostly restricted by the SN1987a energy-loss arguments
(Keil et al.  1997).

Very  recently,  Dom\'\i  nguez,  Straniero  \& Isern  (1999)  have used
Asymptotic   Giant  Branch  (AGB)  stars  as   promising   astroparticle
laboratories.  They concluded  that the  characteristics  of the thermal
pulses  ensuing after the  exhaustion  of He at the core are modified by
the inclusion of axion  emission and that the mass of the  carbon-oxygen
degenerate  core is much lower when axion losses are taken into account.
AGB stars  are  supposed  to be the  progenitors  of white  dwarfs  and,
therefore,  white dwarfs are as well  excellent  candidates  to test the
existence of several  weakly  interacting  massive  particles.  This was
early  recognized by several  authors (see, for instance  Raffelt 1996).
There is, however, another  important reason for this:  white dwarfs are
well  studied  stellar  objects  since the  relative  simplicity  of the
physics   governing  their   evolution  at  moderately   high  effective
temperatures  (say $T_{\rm eff}  \gtrsim  6000$ K) allows to obtain very
detailed models which can be satisfactorily compared with their observed
properties.  Among  white  dwarfs  there is a  specific  class of stars,
known as ZZ-Ceti  objects,  which have a  hydrogen-rich  envelope  (thus
being  classified as DA white dwarfs) and show  periodic  variations  in
their light curves.  G117-B15A  belongs to this particular set of stars.
The observed  periods of  pulsation  are 215.2, 271 and 304.4 s together
with harmonics and linear  combinations of the quoted periods, being the
dominant  pulsation  mode the 215.2 s mode.  The  luminosity  variations
have  been  successfully   explained  as  due  to  $g$-mode  pulsations.
G117-B15A has been recently  claimed to be the most stable optical clock
ever found,  being the rate of change of the 215.2 s period  very small:
$\dot{P}=  (2.3 \pm 1.4)  \times  10^{-15}  {\rm  s\,  s}^{-1}$,  with a
stability  comparable  to that of the most  stable  millisecond  pulsars
(Kepler  et al.  2000).  The rate of change  of the  period  is  closely
related to its cooling timescale, which can be accurately computed, thus
offering a unique  opportunity to test any additional (or  hypothetical)
sink of energy.  This fact was first  recognized  by Isern,  Hernanz  \&
Garc\'\i  a--Berro (1992), who derived an upper bound to the mass of the
axion of 8.7 meV (assuming  $\cos^2\beta=1$) by using a simplified model
and comparing the observed and computed  rate of change of the period of
the 215.2 s mode.

In this paper we use the observational  characteristics of the pulsating
white dwarf  G117-B15A to obtain an upper limit to the mass of the axion
and  we  show  that  pulsating  white  dwarfs  are  powerful  tools  for
constraining  the  mass of any  hypothetical  elementary  particle.  The
paper is  organized  as follows.  In  \S\ref{sec_inputs}  we explain our
input physics and how the models are  computed.  In  \S\ref{sec_obs}  we
briefly  summarize  the  observational  properties  of G117-B15A  and we
discuss the observational  uncertainties of the relevant parameters.  In
\S\ref{sec_modes}   we  extensively   discuss  the  mode  identification
procedure  and  the  theoretical  uncertainties  due to  our  incomplete
knowledge of the adopted  input  physics and we compare our results with
those of other  authors.  In section  \S\ref{sec_axion}  we compute  the
effects of introducing the axion emissivity in our  calculations  and we
use the observed  properties of G117-B15A to impose  constraints  to the
mass of the  axion.  Finally  in  \S\ref{sec:conclu}  we  summarize  our
conclusions.


\section{Our    input   physics   and   the   method   of   calculation}
\label{sec_inputs}

In order to compute the $g$-modes of the white dwarf models described in
this paper, we have used our evolutionary-pulsational  code as described
in C\'orsico \& Benvenuto  (2001).  Briefly, both the  evolutionary  and
the pulsational codes are written following a finite differences  scheme
and the  solutions  are reached by means of  Newton-Raphson  iterations;
such a method is well known in stellar  astrophysics  and it is  usually
referred to as the Henyey method.  The  evolutionary  code contains very
detailed physical  ingredients.  Among others perhaps the most important
are:  an updated version of the equation of state of Magni \& Mazzitelli
(1979), and the radiative OPAL opacities  (see Rogers \& Iglesias  1998,
for an excellent  review).  Conductive  opacities,  neutrino emission by
several processes, and other physical  ingredients are included as well.
For more details, the reader is referred to Benvenuto \& Althaus  (1998)
and references  therein.  On its hand, the pulsational  part of the code
solves the equations for linear  non-radial  stellar  pulsations  in the
adiabatic approximation (Unno et al.  1989).

Briefly,  non-radial  $g$-modes are a subclass of spheroidal modes whose
main restoring force is gravity.  These modes are  characterized  by low
oscillation  frequencies  (long  periods) and by a  displacement  of the
stellar  fluid   essentially   in  the   horizontal   direction.  For  a
spherically  symmetric star in the linear  approximation, a $g$-mode can
be   represented   as  a  standing  wave  of  the  form   $f^\prime_{\rm
k,l,m}(r,\theta,\phi,t)=   g^{\prime}_{\rm  k,l,m}(r)\;  Y^{\rm  m}_{\rm
l}(\theta,\phi)\;  {\rm e}^{i  \sigma_{\rm  k,l,m} t}$, where the symbol
``$^\prime$''  indicates  a  small  Eulerian  perturbation  of  a  given
quantity  $f$  (like  the  pressure).  On their  hand,  $Y^{\rm  m}_{\rm
l}(\theta,\phi)$ are the corresponding spherical harmonics.  Physically,
$l$ is the number of nodal  lines in the stellar  surface and $m$ is the
number of such nodal  lines in  longitude.  In absence  of any  physical
agent  able to  remove  spherical  symmetry  (like  magnetic  fields  or
rotation), the eigenfrequencies  $\sigma_{\rm k,l,m}$ are dependent upon
$l$ but are $2l+1$  times  degenerate  in $m$.  Finally,  $g^\prime_{\rm
k,l,m}(r)$ is the radial part of the eigenfunctions, which for realistic
models must be computed numerically together with $\sigma_{\rm  k,l,m}$.
The index $k$ (known as the radial order of the mode) represents, in the
frame of simple stellar models (like those of white dwarf stars which we
shall study  below), the number of nodes in the radial  component of the
eigenfunction.  For $g$-modes, the larger the value of $k$ is the longer
the  oscillation  period.  For  a  detailed  description  of  non-radial
oscillations in stars, see the textbook by Unno et al.  (1989).

Our aim is to compute  the  eigenmodes  and  particularly  the values of
$\sigma_{\rm  k,l,m}$  during the evolution of a given  stellar  object.
Now, let us  describe  how our  pulsation  and  evolutionary  codes work
together.  Firstly,  an  interval  in  effective  temperature  is chosen
(hereafter  referred  to as the  $T_{\rm  eff}$-strip),  as well  as the
frequency  window  to  be  scanned.  The   evolutionary   code  computes
self-consistently  the white  dwarf  cooling  up to the  moment  when it
reaches  the blue  (hot)  edge of the  $T_{\rm  eff}$-strip.  Then,  the
program calls the pulsation routine and begins to scan for the pulsating
modes.  When a mode is found, the code generates an approximate solution
which is improved  iteratively.  Then, such  solution is tested  and, if
necessary,  the  meshpoint  distribution  is refined.  The mode is again
iterated until satisfactory convergence is reached and then it is stored
with the aim of being employed later as an approximate  solution for the
next stellar model of the  sequence.  This  procedure is repeated  until
the relevant  frequency  window is fully covered.  At this point we have
finished the  computation of all the modes of the first model  belonging
to the $T_{\rm  eff}$-strip  and the structure of each one is now in the
computer memory.  Then, the evolutionary code generates the next stellar
model  and  the  code  calls  the  pulsation  routine  again.  Now,  the
previously stored modes are taken as an input to the iterative scheme to
approximate  the  modes of this  subsequent  stellar  model.  Again  the
solution is iterated, and so on.  The whole  procedure is  automatically
repeated for all the evolutionary models inside the $T_{\rm eff}$-strip.
When the model star evolves out the $T_{\rm eff}$-strip, the calculation
is finished.

An  important  quantity  needed to compute the  pulsational  spectrum of
$g$-modes  is the  so-called  Brunt-V\"ais\"al\"a  frequency,  since  it
basically defines the main pulsational  properties of white dwarfs.  Due
to the  compositional  changes along the white dwarf structure, there is
an  important  local  contribution  to this  frequency  at the  chemical
interphases,  which, in turn, is the  responsible  for the occurrence of
mode trapping (that is, modes with a non-negligible  amplitude which are
confined  between two different  chemical  interphases or one interphase
and the surface).  In particular,  for the case of DA white dwarfs, mode
trapping is mainly due to the H-He interphase.  The  Brunt-V\"ais\"al\"a
frequency is usually given by (Unno, et al.  1989):

\begin{equation}
N^{2} = g\ \left(\frac{1}{\Gamma_{1}}\ \frac{d\ln P}{dr} - 
\frac{d\ln \rho}{dr} \right)
\end{equation}

\noindent  where all the  symbols  have  their  usual  meaning  (notice,
however, that in this  paragraph $P$ stands for the pressure  whereas in
the rest of the paper it  stands  for the  period  of  oscillation)  and
$\Gamma_1$  is the first  adiabatic  exponent.  However, and in order to
avoid  numerical  noise, we use the expression  given by Brassard et al.
(1991):

\begin{equation}
N^2   =   \frac{g^2\   \rho}{P}\   \frac{\chi_{\rm   T}}{\chi_{\rho}}\
\left(\nabla_{\rm ad} - \nabla + B \right).
\end{equation}

\noindent where $\chi$ denotes the partial logarithmic derivative of the
pressure  with  respect to either $T$ or $\rho$,  respectively,  and $B$
stands for

\begin{equation}
B=-\frac{1}{\chi_{\rm T}} \sum^{N-1}_{i=1} \chi_{\rm X_{\rm i}} 
\frac{d\ln {X}_{\rm i}}{d\ln P} 
\end{equation}

\noindent  where  $X_{\rm i}$ is the mass  fraction  of atoms of species
$i$, $N$ is the total number of considered species, and

\begin{equation} 
\chi_{\rm X_{\rm i}}= \left( \frac{\partial \ln{P}}
{\partial \ln{X_{\rm i}}} \right)_{\rho,T,\{X_{\rm j \neq i}\} }.
\end{equation}

Although  we do not mean to  discuss  in  depth  the  comparison  of our
results with those obtained with other different  numerical codes, it is
noteworthy  that  we  have   thouroughly   tested  the  results  of  the
pulsational part of the code by comparing our numerical  results for the
dimensionless  eigenvalues  with  those  of  the  polytropic  models  of
Christensen--Dalsgaard   \&  Mullan   (1994),   and  with  the   periods
corresponding to DA white dwarf models kindly provided by Bradley (2000,
private  communication).  To be specific, we have tested our pulsational
code with two  carbon-oxygen  DA white dwarf models of 0.50  $M_{\odot}$
and 0.85  $M_{\odot}$, the structure of which was computed with the WDEC
evolutionary  code  and  its  vibrational   properties  were  previously
analyzed (Bradley 1996).  We considered a large number of modes, and the
differences  between our  computed  periods and those of Bradley  (2000)
were always  smaller than  $\approx 0.1 \%$.  It is also worth  noticing
that we also tested the  influence  of the number of  meshpoints  in the
computed  eigenvalues.  We have found that largely increasing the number
of meshpoints  in the  pulsational  calculation  does not  substantially
affect the computed  periods.  In fact the values  changed very slightly
($\ll 0.1 \%$).

In order to compute the effects of axion emission, we have  incorporated
into the evolutionary code the axion emission rates of Nakagawa, Kohyama
\&  Itoh  (1988).  Axion  emission   produces  a  supplementary   energy
loss-rate to those that are  considered in the standard  theory of white
dwarf evolution and, consequently, axions accelerate the cooling process
of white dwarfs.  Such an  acceleration  of the  evolution  has a direct
consequence  on the  pulsational  properties  of the object,  because it
produces an enhanced value of the period  derivative of the  oscillation
modes,  $\dot{P}$  (Isern,  Hernanz  \&  Garc\'\i  a--Berro,  1992).  Of
course, due to this extra cooling  mechanism, the structure of the white
dwarf  itself is also  affected but in such a way that, as we shall show
below, for a given $T_{\rm eff}$ it  corresponds  a period, $P$, that is
largely independent of the exact value of the axion energy losses.  This
fortunate  fact will allow us to  identify  an  internal  structure  for
G117-B15A  independently  of the axion  emission  rates,  which  largely
simplifies our analysis.


\section{Some     observational     characteristics     of    G117-B15A}
\label{sec_obs}

G117-B15A is an otherwise  typical DA  (hydrogen-rich)  white dwarf star
whose  variability  was first  discovered by McGraw \& Robinson  (1976).
Since then on, it has been monitored almost  continuously.  The mass and
effective temperature of this star have been spectroscopically estimated
to be 0.59  $M_{\odot}$  and  11,620 K,  respectively  (Bergeron  et al.
1995).  More  recently,  Koester  \&  Allard  (2000)  have  suggested  a
somewhat lower value for the mass, which appears to be 0.53 $M_{\odot}$.

Regarding the  variability  of this star, its observed  periods are (see
Kepler et al.  1982) 215.2, 271 and 304.4 s together with  harmonics and
linear  combinations of the quoted periods.  Of particular  interest for
this work is the fact that for the 215.2 s mode it has been  possible to
find its  rate of  change,  $\dot{P}$.  The  first  published  value  of
$\dot{P}$  for this star was  presented by Kepler et al.  (1991).  These
authors derived an {\sl upper limit} to the rate of change of the period
using  all the  data  obtained  from  1975 to  1990.  This  upper  limit
appeared  to be  $\dot{P}=  (12.0  \pm 3.5)  \times  10^{-15}  {\rm  s\,
s}^{-1}$,  much  larger  than the  theoretical  predictions.  Along this
decade  several  upper  limits to the rate of change of the 215.2 s mode
have been  derived,  with  values  decreasing  from a  maximum  value of
$\dot{P}=  (3.2 \pm 3.0) \times  10^{-15} {\rm s\, s}^{-1}$ to a minimum
value of $\dot{P}=  (1.2 \pm 2.9) \times  10^{-15} {\rm s\, s}^{-1}$ ---
see Kepler et al.  (2000),  for a detailed  discussion  of the  measured
values of $\dot{P}$.  Very recently, with a much longer time interval of
acquired data, Kepler et al.  (2000)  re-determined  $\dot{P}$ finding a
significantly  lower value of  $\dot{P}=  (2.3 \pm 1.4) \times  10^{-15}
{\rm  s\,  s}^{-1}$,  and  concluded  that,  in  view  of the  presently
available data, the 215.2 s mode of G117-B15A is perhaps the most stable
oscillation  ever  recorded  in  the  optical  band,  with  a  stability
comparable  to that of the  most  stable  millisecond  pulsars.  To this
regard it is important to realize  that, as already  mentioned,  all the
previous  determinations  of the  rate  of  change  of the  period  were
actually upper limits.  The reason for this improved accuracy is that as
time (squared) passes by the baseline is much larger and,  consequently,
the error bars  decrease,  leading  to more  precise  measurements.  The
important  fact  however  is that all the  previous  upper  limits  were
consistent  each  another  at the  $1\sigma$  level  and  that  the last
measured  value --- that of Kepler et al.  (2000) --- is larger than its
corresponding  uncertainty  yielding,  thus, a  positive  detection  and
providing a precise measurement of $\dot{P}$.

An  important  aspect  for  the  present  work  is  the  so-called  mode
identification.  That is, the  identification  of the $l$ and $k$ values
corresponding  to each observed  period.  In the case of G117-B15A,  the
results  of  Robinson  et al.  (1995)  indicate  that the 215.2 s period
corresponds  to a dipolar mode  ($l=1$).  Following  the work by Bradley
(1998) we shall  assume  that the other two modes  cited  above are also
dipolar  (Brassard  et al.  1993;  Fontaine \& Brassard  1994), and also
that the other periodicities  present in the light curve of the star are
not associated with actual eigenmodes but are actually due to non-linear
effects in the envelope  (Brassard et al.  1993).  Regarding  the radial
order of the modes, there exist two possible identifications.  Following
Clemens (1994) the observed  periods are dipolar modes with $k=2$ (215.2
s), $k=3$ (271 s) and $k=4$  (304.4  s).  On the other  hand,  Fontaine,
Brassard \&  Wessemael  (1994)  identified  these modes as dipolar  with
$k=1$ (215.2 s), $k=2$ (271 s) and $k=3$ (304.4 s).  As we shall show in
the following  section, our best fit to  observations  is in  accordance
with the identifications of Clemens (1994).


\section{The    internal   structure   of   G117-B15A   and   the   mode
identification}\label{sec_modes}

For the above mentioned value of the stellar mass, the theory of stellar
evolution  predicts an interior rich in carbon and oxygen.  The chemical
composition  of the white dwarf  interior as a function  of its mass has
recently been  computed in detail by Salaris et al.  (1997).  Therefore,
for the calculations  reported here we shall adopt the chemical profiles
obtained by these authors.  However it is worth  mentioning  that a good
deal of work  has  been  done to this  regard.  In  particular  Bradley,
Winget \& Wood (1992), among  others, have studied the  influence of the
internal  core  composition  on the  structure of white dwarfs and, more
specifically  in the properties of DA pulsators.  More recently  Bradley
(1996, and references  therein) analyzed several models for DA pulsators
and compared his  calculations  with the  observations  finding that the
chemical  composition  of the  carbon-oxygen  core is more  likely to be
oxygen-rich, in good agreement with the chemical  profiles of Salaris et
al.  (1997).  On its hand,  Dom\'{\i}nguez  et al.  (1999) have computed
updated chemical profiles for the AGB progenitors of carbon-oxygen white
dwarfs, finding chemical stratifications which are very similar to those
of  Salaris  et  al.  (1997).  In  this  work  we  have   considered  as
relatively  reliable  only the chemical  composition  of the core, which
will not be changed unless  explicitly  stated (however, see below).  In
order to  adjust  the  pulsational  properties  of our  models  to those
observed in G117-B15A, we have looked for  variations of some details of
the chemical  profiles and thicknesses of the layers that at present are
only poorly  determined,  like the thickness of the hydrogen  very outer
layer and of the helium layer that is underneath  it.  To be precise, in
looking  for  such a  fitting  of the  oscillatory  properties  we  have
considered  the  thicknesses  of the hydrogen and helium  layers as free
parameters.  There are several  theoretical  reasons for suspecting that
this is a safe procedure, but perhaps the most important one is that the
thickness of the hydrogen layer strongly  depends on the exact moment at
which a white dwarf  progenitor  departs from the thermally  pulsing AGB
phase through pulsation-driven mass-loss (see, for instance, Iben 1984).
Regarding  the  thickness of the helium layer it is important to mention
here that, although the theoretical reasons are not so compelling, it is
a usual  procedure to adopt it as a free  parameter as well, and that as
it occurs with the hydrogen  layer notably  influences  the cooling rate
and, consequently, the rate of change of the period.

Another important  quantity is the slope of the chemical profiles of the
very outer layers, especially in the  helium-hydrogen  and carbon-helium
interphases.  These slopes have a strong  influence  on the  pulsational
properties of white dwarfs and,  especially, in the  Brunt-V\"ais\"al\"a
frequency,  which  ultimately can lead to mode trapping (see Brassard et
al.  1992).  The  functional  dependence  adopted  in this work for both
composition  transition  interphases  is  gaussian.  At present it seems
possible  to  look  for a  more  physically  sound  treatment  of  these
interphase regions since the structure of such zones should be (at least
in the case of those that occur in non strongly  degenerate  conditions)
determined  by  element  diffusion.  In  fact,  diffusion  is  the  main
responsible  for expecting the outer layers to be made up of almost pure
materials and should also  ultimately  determine  the  thickness of such
layers.  Some authors  have  incorporated  the effects of  diffusion  in
computing white dwarf evolution (see Iben \& MacDonald 1985, and Althaus
\& Benvenuto  2000) but, to our knowledge, no one has considered yet the
effects of diffusion on the pulsational  properties of evolving DA white
dwarfs.  We thus defer such a study to a forthcoming publication.


\begin{figure} 
\centering
\vspace*{14cm}
\includegraphics{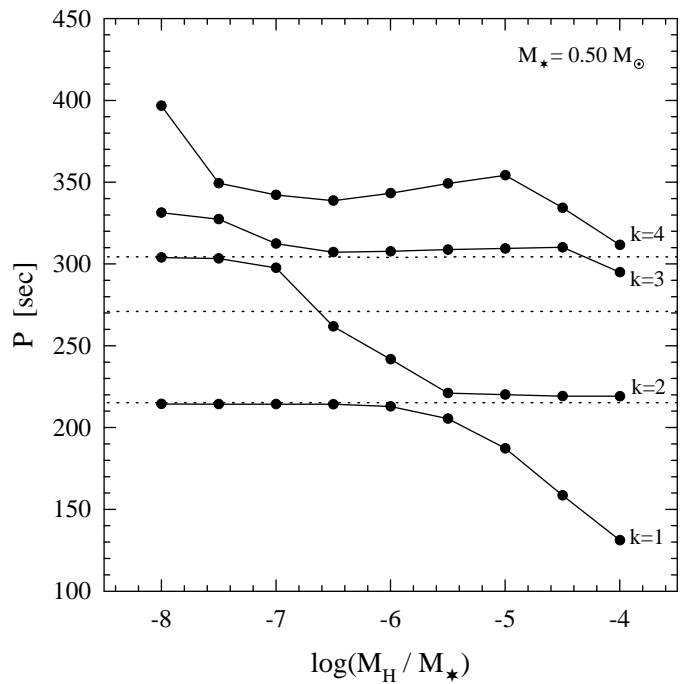}
\caption{The  period of the dipolar modes $k=1,\, 2,\, 3$, and $4$ for a
0.50  $M_{\odot}$  model at $T_{\rm eff}= 11,620$ K as a function of the
mass fraction in the outermost hydrogen layer.  The rest of the chemical
profiles  correspond  to the best  obtained fit (see text for  details).
The filled dots  correspond  to the computed  models,  whereas the short
dashed  lines  represent  the  observed  oscillation  periods  found  in
G117-B15A.  In this  case of a rather  low  stellar  mass,  the best fit
corresponds   to  the   modes   $k=1$,   2,  and  3  with   $\log{M_{\rm
H}/M_{\star}}=  -6.6$.  For the rest of the computed  cases, neither the
215.2 s period nor the two other are well fitted.}
\label{fig:0.50}
\end{figure}

For our  purposes, we have looked for a model with the observed  $T_{\rm
eff}$ and mass for  G117-B15A  which {\sl  simultaneously}  matches  the
three observed  modes as good as possible.  After having such a fiducial
model, the computation of the theoretical $\dot{P}$ for different values
of the  axion  mass  is  rather  straightforward.  In  looking  for  the
fiducial  model, we have handled some  characteristics  of the  chemical
profiles as free (around the values  predicted by the standard theory of
stellar  evolution)  and computed the evolution of the  resulting  model
down to the effective  temperatures  corresponding to that of G117-B15A.
Then, we  computed  the modes to be  compared  with the  observed  ones.
Given the present uncertainties, the mass of the star and the helium and
hydrogen  mass  fractions  were  considered  to  lay  in  the  following
intervals:  $0.50  \leq   M_{\star}/M_{\odot}   \leq   0.65$,  $-3  \leq
\log{M_{\rm He}/M_\star} \leq -2$, $-8 \leq \log{M_{\rm H}/M_\star} \leq
-4$.  Also, as quoted above, we considered different thicknesses for the
helium-hydrogen and carbon-helium  interphases.  Defining the lagrangian
mass-coordinate as $q= \log{(1-M_{\rm  r}/M_\star)}$,  and $\Delta q$ as
the thickness of the interphase  layer, we have considered the following
reasonable  values:  $-0.8 \leq \Delta q_{\rm H-He} \leq -0.4$ and $-0.8
\leq \Delta  q_{\rm  He-C} \leq  -0.2$.  As it will be clear  below, the
simultaneous fit of the three modes is a strong  condition to be imposed
to the slopes of the chemical profile of the interfaces of the star.


\begin{figure} 
\centering
\vspace*{14cm}
\includegraphics{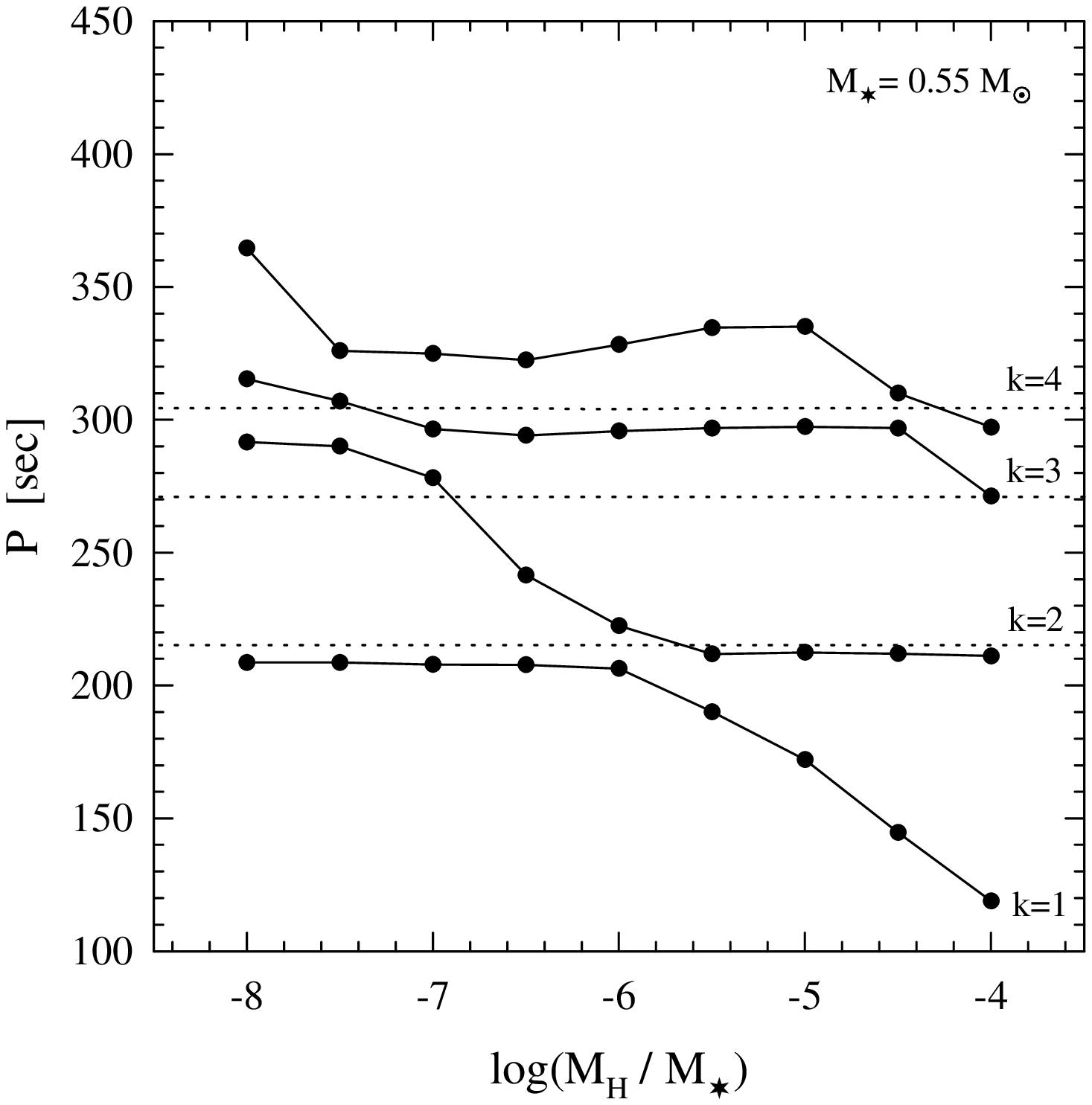}
\caption{Same  as in Fig.  \ref{fig:0.50}, but for  the  case  of a 0.55
$M_{\odot}$  model.  In this case, the best fit to the  observations  is
found for $k=2$, 3, 4 and $\log{M_{\rm  H}/M_\star}=  -4.0$.  As this is
the best fit to the observations we have found so far, we shall refer to
it as the fiducial model.  Notice that the amount of hydrogen present in
the fiducial model is in very nice agreement  with the  expectations  of
the standard  theory of stellar  evolution.  Another  acceptable  fit is
obtained with $k=1$, 2, 3 and  $\log{M_{\rm  H}/M_\star}=  -7.0$ but the
differences between the observed and computed values are far larger than
in the previously mentioned case.}  
\label{fig:0.55}
\end{figure}

\begin{figure} 
\centering
\vspace*{14cm}
\includegraphics{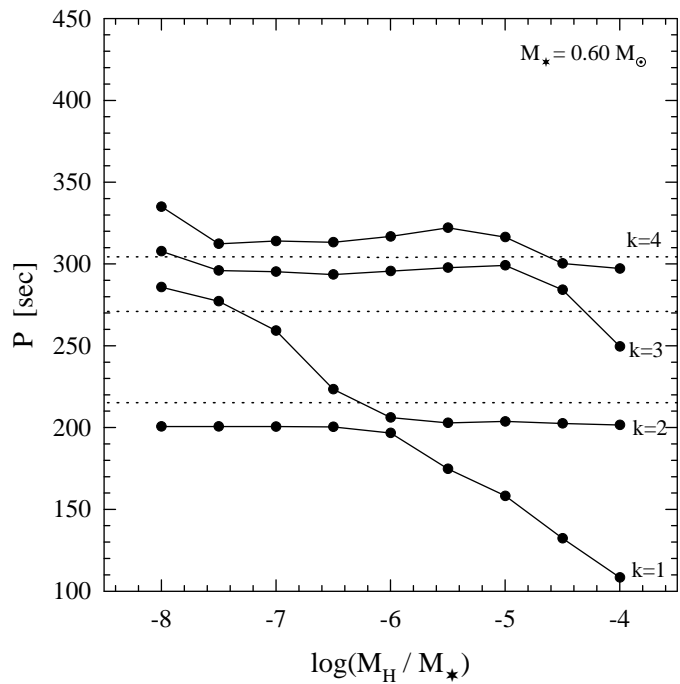}
\caption{Same  as in Fig.  \ref{fig:0.50},  but for the  case  of a 0.60
$M_{\odot}$ model. In this case, we have found that the observations are
not well fitted by any  of the considered hydrogen mass-fractions.  This
strongly suggests, in accordance with the findings of other authors (see
text),  that the mass of this  white  dwarf  is  lower  than  the  value
considered in this figure.}
\label{fig:0.60}
\end{figure}

In  Figs.  \ref{fig:0.50}-\ref{fig:0.60}   we  compare  the  theoretical
periods  of the $l=1$  (dipolar)  modes  with  $k=1, 2, 3,$ and $4$ with
those  observed for  G117-B15A.  In  constructing  these figures we have
considered  models  with the  observed  $T_{\rm  eff}$ of 11,620  K, and
previously  adjusted  the rest of the  characteristics  of the  chemical
profile to our fiducial values ($\log{M_{\rm  He}/M_\star}= -2$, $\Delta
q_{\rm H-He}= -0.8$, and $\Delta  q_{\rm  He-C}=  -0.4$)  except for the
most  critical  one, which turns out to be the  fraction  of the stellar
mass  present  in the  outermost,  pure  hydrogen  layer.  It  should be
mentioned  as well that in  computing  these fits we have not taken into
account  the  axion  emissivity.  In the  case of the  0.50  $M_{\odot}$
models, it is found a rather good fit  corresponding to the modes $k=1$,
2, and 3 with  $\log{M_{\rm  H}/M_\star}=  -6.6$.  In the case of a 0.55
$M_{\odot}$  model  star we find  that the best fit  corresponds  to the
modes with  $k=2$, 3, and 4 and  $\log{M_{\rm  H}/M_\star}=  -4.0$.  Yet
there is another  acceptable  fit for  $k=1$, 2, and 3 and  $\log{M_{\rm
H}/M_\star}= -7.0$, although the previously mentioned fit can undoubtely
considered as much better.  In the case of the 0.60  $M_{\odot}$  models
we have found no  satisfactory  fit for any of the  considered  hydrogen
mass  fractions.  Let us also  remark  at this  point  that we have also
considered the case of models with 0.65 $M_{\odot}$, which are not shown
for the  sake of  conciseness.  


As explained above, another important free parameter is the thickness of
He-rich  layer,  since  it  also   influences  to  a  large  extent  the
identification  of modes and the  cooling  rate.  To this  regard  it is
important  to mention  that we have also  explored  the  possibility  of
thinner helium-rich layers  ($\log{M_{\rm  He}/M_\star} < -3.0$) for all
the mass values quoted above.  Thicker helium layers are quite  unlikely
according to the most recent AGB studies  (Dom\'\i  nguez et al.  1999).
In all of these  cases  the fits to the  observations  turned  out to be
significantly  worse.  Thus we have discarded them and, consequently, we
do not show the  corresponding  fits.  Note as well  that  besides  this
being in  accordance  with the  predictions  of the  theory  of  stellar
evolution it is also in agreement with the results of Bradley (1998).


\begin{figure}   
\centering   
\vspace*{14cm}    
\includegraphics{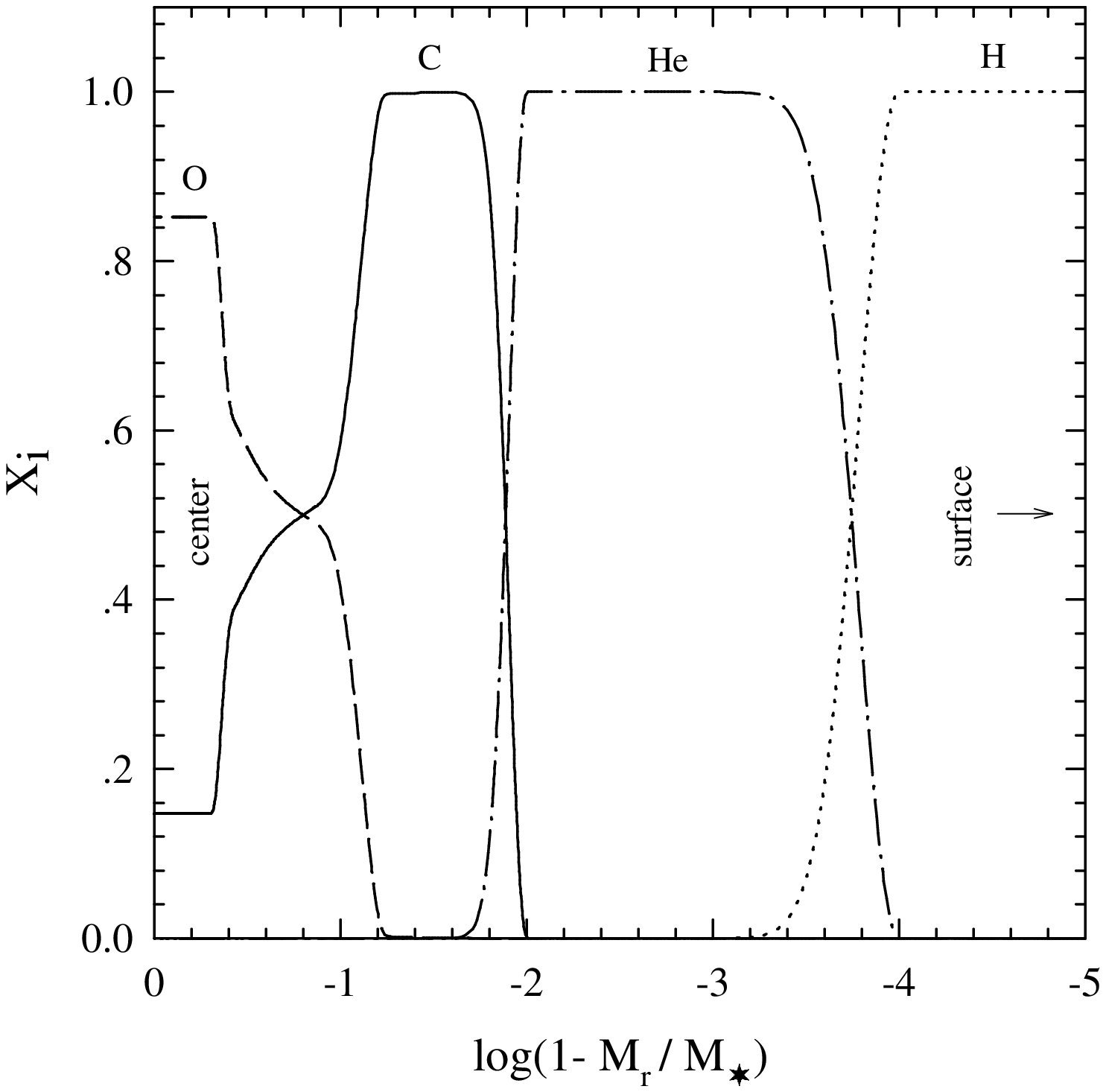}
\caption{The internal chemical composition of the white dwarf model that
best  matches  the  observed  pulsation  periods of  G117-B15A.  This is
essentially  the profile  obtained  by Salaris et al.  (1997) for a mass
value of 0.55  $M_{\odot}$.  The  composition of the white dwarf core is
$\approx 83\%$ by mass of oxygen, being the remaining 17\% carbon.  Atop
the degenerate  core there is a helium layer of about $1\%$ of the total
mass.  The outermost hydrogen layer embraces $10^{-4} \, M_\star$, which
is in nice  agreement  with the  predictions  of the  stellar  evolution
theory.  Notice  that only the  interphases  between  carbon-helium  and
helium-hydrogen   have  been  tuned  to  fit  the  observations.  For  a
discussion of the physical plausibility of such tuning, see text.}
\label{fig:fiducial} 
\end{figure}


\begin{figure} 
\centering
\vspace*{14cm}
\includegraphics{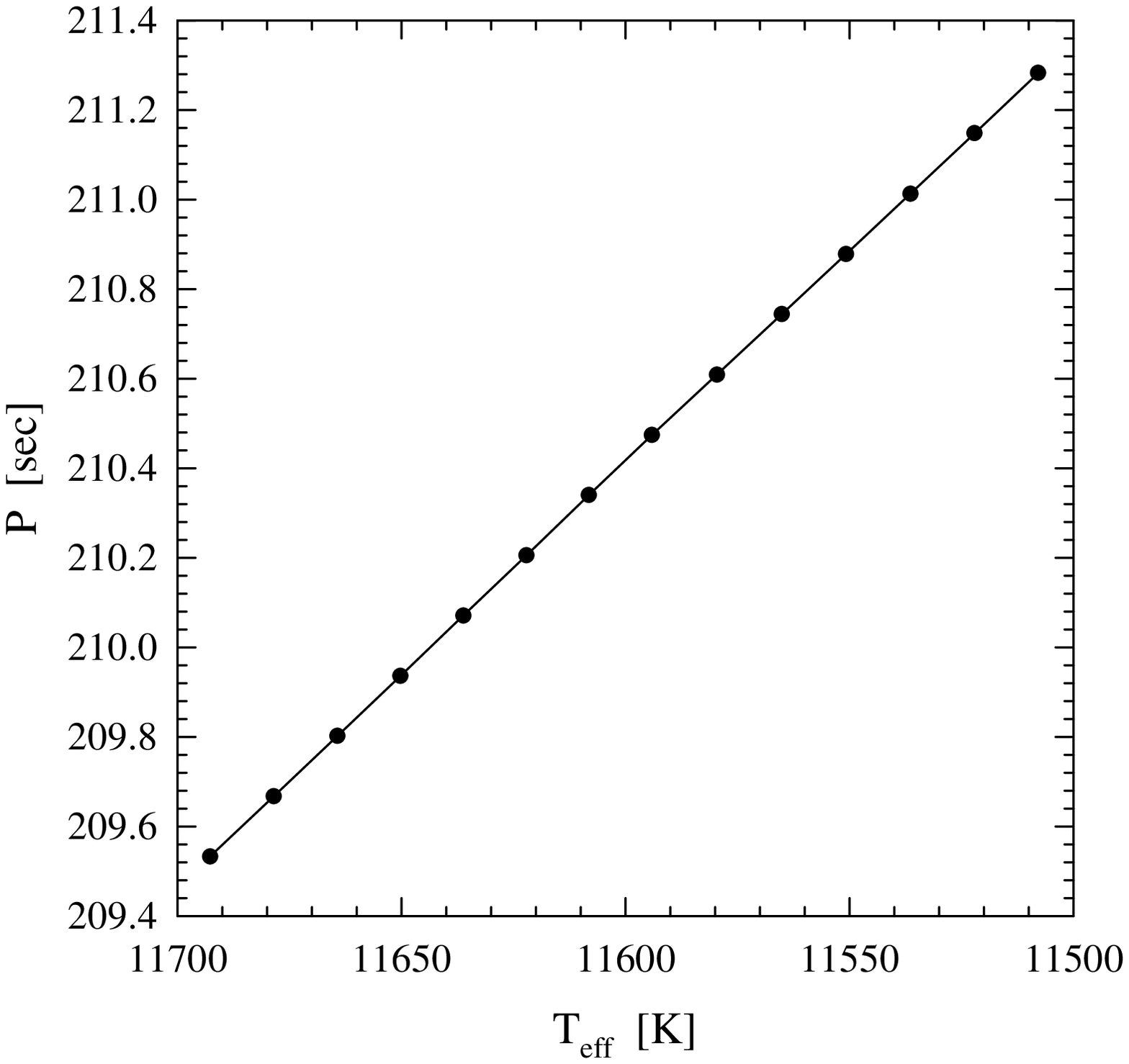}
\caption{The  period of the $l=1$, $k=2$ mode of the fiducial model as a
function of $T_{\rm  eff}$.  The $T_{\rm eff}$  interval has been chosen
to be the  corresponding  to G117-B15A $\pm 200$ K.  Filled dots are the
computed  models.  In this case we have not considered  axion  emission.
Note the monotonic increase in the  exact value of the period due to the
decreasing temperature of the interior.}
\label{fig:p-teff-fiducial}
\end{figure}

These results clearly indicate that the mass of G117-B15A should be very
close to 0.55 $M_{\odot}$ and that the hydrogen mass fraction present in
the star should also be close to $M_{\rm H}/M_\star= 10^{-4}$.  In fact,
the value we derive for the mass of G117-B15A is nicely bracketed by the
independent spectroscopic  determinations of Bergeron et al.  (1995), who
obtained 0.59 $M_{\odot}$, and of Koester \& Allard (2000), who obtained
0.53  $M_{\odot}$.  It is also worth  noticing that our best fit is very
similar to the one presented by Bradley (1998)  assuming an $l=1$, $k=2$
mode for the 215.2 s period, although this author used a stellar mass of
$0.60 M_{\odot}$.  Notably, the amount of hydrogen that we have found in
G117-B15A is in very nice agreement with the predictions of the standard
theory of stellar  evolution.  The model that  provides  the best fit to
the  observations   ($M_\star=   0.55\,   M_\odot$,   $k=2$,  3,  4  and
$\log{M_{\rm  H}/M_\star}=  -4.0$) will be hereafter  referred to as the
fiducial model.

In Fig.  \ref{fig:fiducial} we show the profiles of chemical composition
of the fiducial  model.  These  profiles are  essentially  the  profiles
presented by Salaris et al.  (1997)  corresponding  to the derived  mass
value of $\sim 0.55\,  M_{\odot}$.  The abundance (by mass  fraction) of
oxygen in the degenerate  core is $\approx  83\%$.  Atop the  degenerate
core there is a partially  degenerate helium layer of about $1\%$ of the
total mass.  Finally,  as quoted  above, the  outermost  hydrogen  layer
embraces  $10^{-4}\,  M_\star$.  It is interesting  to note as well that
most of the mass of the white dwarf is in the form of oxygen.


\begin{figure} 
\centering
\vspace*{14cm}
\includegraphics{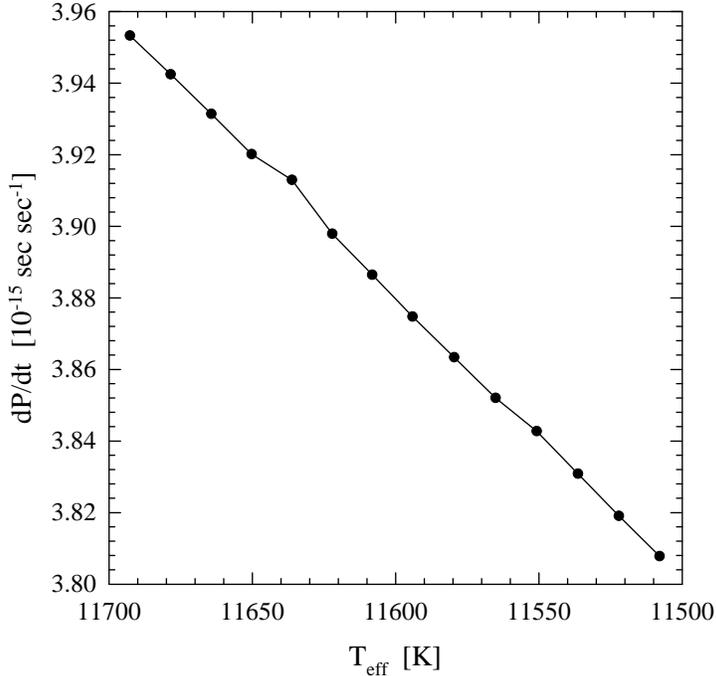}
\caption{Same  as Fig.  \ref{fig:p-teff-fiducial}  but for the  temporal
derivative of the period  corresponding to the $l=1$, $k=2$ mode.  As in
the previous figure we have not considered  axion emission.  Thus, these
results are a lower  theoretical  limit for the temporal  derivative  of
this mode.}  
\label{fig:pdot-teff-fiducial}
\end{figure}

Another  relevant  issue is the  precision of the computed  periods.  In
other words, which are the typical errors of the theoretical  periods of
fiducial model when compared to the observational data of G117-B15A?  We
do not intend here to make sophisticated  comparisons, which in fact are
possible, but instead,  given the discrete  nature of our model grid and
the small  number of periods we are trying to match, we will  perform an
approximate test.  The differences between the observed  frequencies and
the  theoretical  ones are 4.63~s (for  $k=2$),  0.58~s (for  $k=3$) and
7.26~s (for  $k=4$),  thus giving an average  difference  of $\sim 4$~s,
which is in good agreement with the results of Bradley  (1998).  This is
a reasonable  value, since many of the subtle  details of the  pulsation
calculations  can influence the computed  periods at the several seconds
level.  There is still another way of  estimating  the  precision of the
theoretical periods.  Given that we have found that the thickness of the
helium layer largely  influences the computed  periods and given that we
do not get an acceptable  fit for  thicknesses  smaller than  $10^{-2}\,
M_\star$  we will  focus  only  on the  value  of the  thickness  of the
outermost hydrogen layer. We define the quantity

\begin{equation}
\psi=\log \sum_{\rm i=1,2,3} \left( 1- P_{\rm i}^{\rm theor} 
/ P_{\rm i}^{\rm obs}
\right)^{2}
\end{equation}

\noindent and we minimize it as a function of $M_{\rm H}$ and $M_\star$.
We obtain two local minima:  one at $\log(M_{\rm  H}/M_\star)=-4.04$ and
$M_\star   =0.55\,    M_\odot$   and   another   one   at   $\log(M_{\rm
H}/M_\star)=-6.60$  and $M_\star  =0.50\,  M_\odot$.  We have chosen the
first fit because the mass of the  fiducial  model is much closer to the
spectroscopic mass of G117-B15A.  Then we look for the deviations around
this minimum and we find that the typical width of the minimum is of the
order of 2~s.  Since this value is smaller than the previously  obtained
average  value  of the  differences  between  the  theoretical  and  the
observational  data we consider that a reasonable  value of the goodness
of the the fit is  $\simeq  5$~s,  which is the value  quoted in Table 1
below.

Now,  we are in a  good  position  to  study  some  general  vibrational
properties  of the  fiducial  model.  In  particular  we will  study the
dependence of our results on the observational properties (basically the
effective  temperature  and the mass) of  G117-B15A.  This is  important
since the  observational  uncertainties  affect the way the  theoretical
stellar models are compared with the observations  and, thus, can affect
the upper  limit on the mass of the axion  which will be derived  below.
Therefore, we shall explore the effects resulting from varying the above
mentioned  characteristics  of the model, keeping  constant the fiducial
chemical profile  discussed  earlier.  As the mode to be employed is the
$l=1$, $k=2$, we shall hereafter restrict our analysis to such mode.  In
order to do so, in Fig.  \ref{fig:p-teff-fiducial} we show the period of
the $l=1$, $k=2$ mode that best represents the observed  period of 215.2
s as a function of the  effective  temperature  in an interval  centered
around  11,600 K (the  observed  value  for  G117-B15A).  Note  that the
monotonic increase in the period of this mode is a direct consequence of
the cooling process.  On its hand, in Fig.  \ref{fig:pdot-teff-fiducial}
we show the  temporal  derivative  of the  period  corresponding  to the
$l=1$, $k=2$ mode.  It is very important to stress at this point that as
in this  figure  we  have  not  considered  yet  the  possibility  of an
additional axion cooling  mechanism,  these results  represent the lower
theoretical  limit for the $\dot{P}$ of this mode.  As it can be seen in
this  figures  the  errors  in  the   determination   of  the  effective
temperatures   do  not  induce   large   errors   neither  in  the  mode
identification nor in the computed value of $\dot{P}$.


\begin{figure}  
\centering
\vspace*{14cm}
\includegraphics{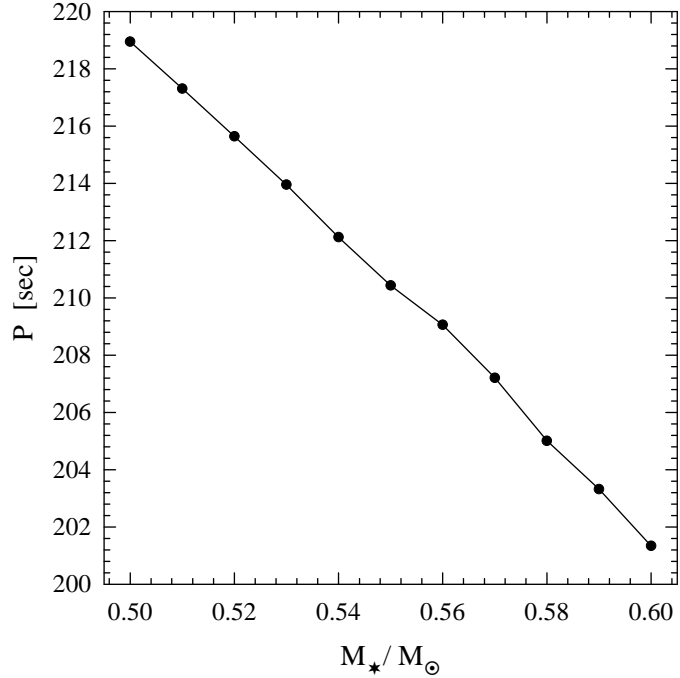}
\caption{The  $l=1$,  $k=2$  period  for  white  dwarf  models  with the
chemical  profile  shown in Fig.  \ref{fig:fiducial}  and $T_{\rm  eff}=
11,620$ K as a function of their mass.  The computed models are depicted
with filled dots.}  
\label{fig:p-m}
\end{figure}


\begin{figure} 
\centering
\vspace*{14cm}
\includegraphics{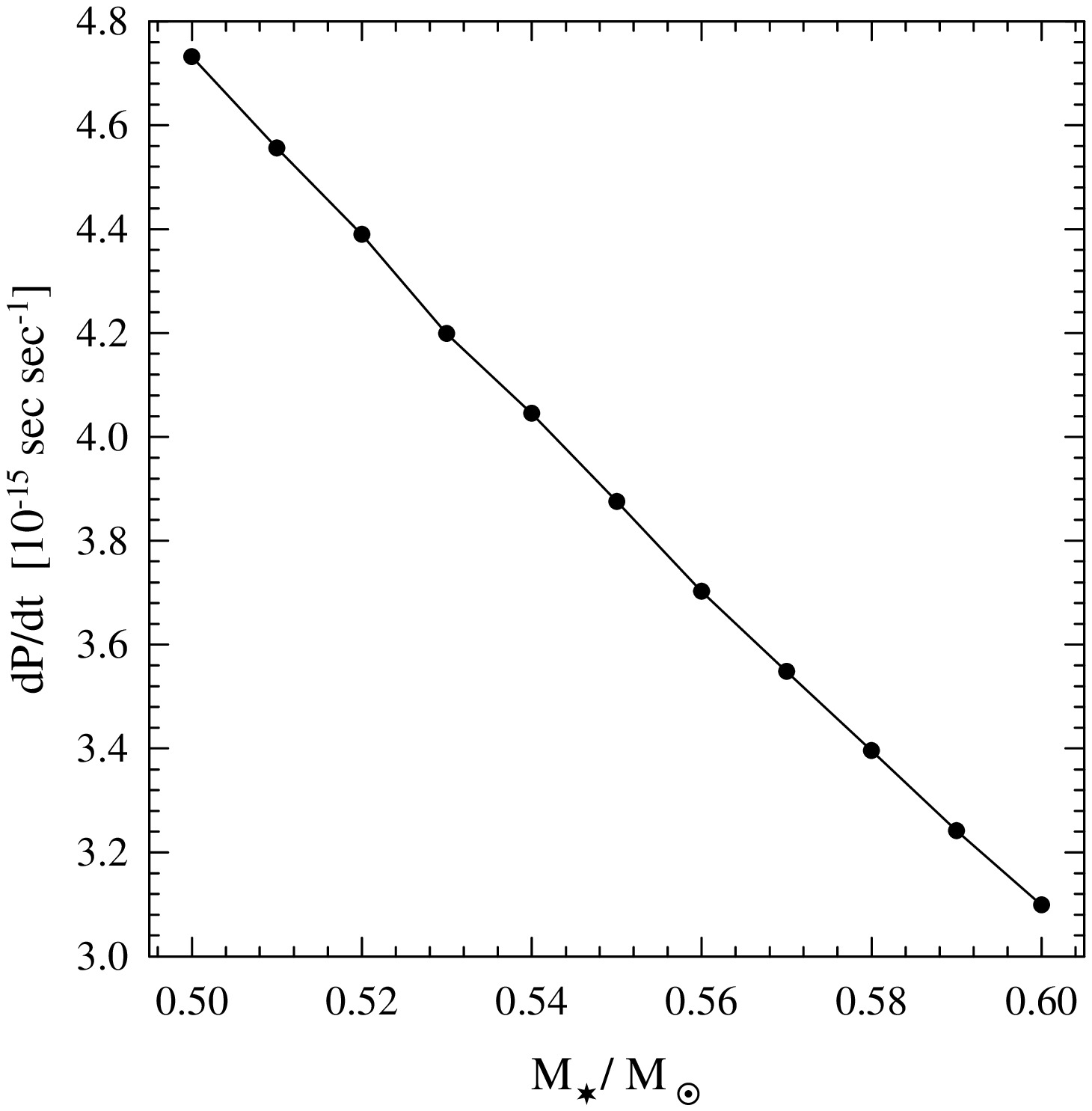}
\caption{The  temporal derivative of the period of the $l=1$, $k=2$ mode
for  white  dwarf  models  with  the  chemical  profile  shown  in  Fig.
\ref{fig:fiducial}  and $T_{\rm  eff}=  11,620$ K as a function of their
mass.  The   computed   models   are   depicted   with   filled   dots.}
\label{fig:pdot-m}
\end{figure}

Now we consider  white dwarf models with the fiducial  chemical  profile
and the same  $T_{\rm  eff}$ and look for the  effects of  changing  the
stellar  mass.  In  Figs.  \ref{fig:p-m}-\ref{fig:pdot-m}  we  show  the
period and its  temporal  derivative  as a  function  of mass.  When one
considers  models  with  increasing   masses  and,  hence,  with  larger
gravities the Brunt-V\"ais\"al\"a  frequency  correspondingly  increases
(see   equation   1).  Since   all   the   eigenfrequencies   ($\sigma$)
corresponding to locally non-evanescent $g$-modes verify  $\sigma^2<N^2$
(see Unno, et al.  1989) it turns out that all the pulsational  spectrum
is shifted to larger  values,  resulting  thus in smaller  periods.  The
same   behaviour   was  found  by  Brassard  et  al.  (1992).  Regarding
$\dot{P}$, the larger the mass, the lower the radiating  surface and the
higher  the total  heat  capacity.  As we are at a fixed  $T_{\rm  eff}$
value, high mass models have smaller  luminosities  and thus cool slower
with a smaller  $\dot{P}$.  This can be also  understood in terms of the
most simple, yet accurate  enough for our purposes,  cooling law (Mestel
1952).  Within  the  framework  of the  Mestel  (1952)  cooling  law the
luminosity of a white dwarf is $L\propto M_\star  T^{7/2}$, where $T$ is
the  temperature of the nearly  isothermal  degenerate  core, which is a
monotonic function of the effective  temperature, and the characteristic
cooling  time  turns  out to be  $\tau_{\rm  cool} = d\ln  L/dt  \propto
(M_\star/L)^{5/7}$.  Since, to an accuracy similar to that of the Mestel
cooling  law, the  secular  rate of change  of the  period  is  directly
related to the cooling rate we expect:

\begin{equation}
\frac{d\ln P}{dt}\propto-\frac{d\ln T}{dt}
                 \propto\frac{1}{\tau_{\rm cool}}
                 \propto\Big(\frac{L}{M_\star}\Big)^{5/7}
                 \propto T^{5/2}.
\end{equation}

\noindent  Since massive  white dwarfs have smaller  radii, for the same
effective  temperature  the  luminosity  is smaller and, hence, the core
temperature is smaller, leading to larger  characteristic  cooling times
and,  ultimately, to smaller secular changes in the period.  In summary,
as in the  previous set of figures we see that the exact value of $P$ is
not   significantly   altered  by  the   observational   errors  in  the
determination  of the stellar  mass.  However,  the  reverse is true for
$\dot{P}$.  In particular it is convenient  to note that the exact value
of the period derivative varies by about a 30\% in the mass interval 0.5
to 0.6  $M_\odot$,  which  can be  important  when  comparing  with  its
observed value.  Nevertheless, and fortunately, the mode  identification
performed  previously  allows  us  to  effectively  constrain  the  mass
interval  to a much  narrower  range and, thus,  this will not  severely
affect our results.


\begin{figure}  
\centering
\vspace*{14cm}
\includegraphics{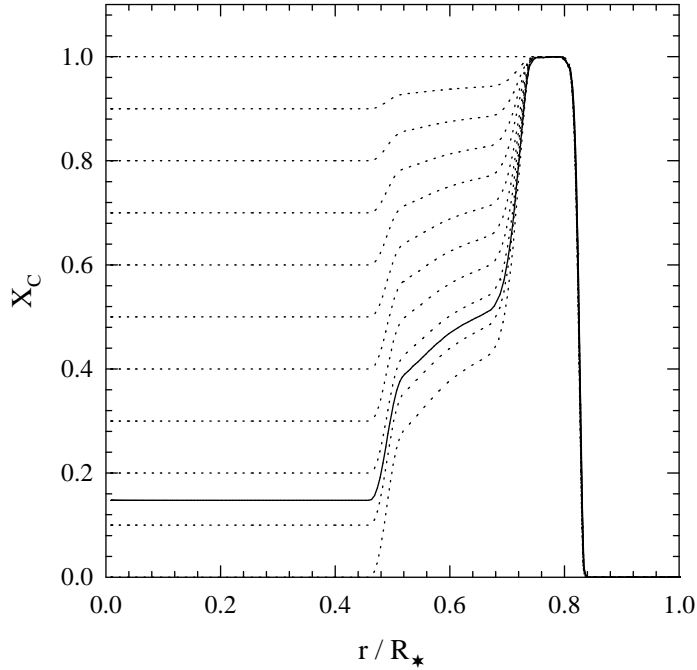}
\caption{Profile  of the carbon  abundance  as a function of the stellar
radius for models with  $X_{\rm  C}=0.0$,  0.1,  $\cdots$,  1.0  central
carbon  abundances.  The short  dashed  lines  correspond  to the scaled
profiles  whereas  the solid  line  correspond  to the  fiducial  carbon
profile.}
\label{fig:prof-scaled} 
\end{figure}

Another possible source of uncertainties  is the poorly  determined rate
of the  reaction  $^{12}$C$(\alpha,\gamma)^{16}$O  that  determines  the
final  state  of the  asymptotic  giant  branch  phase  and,  thus,  the
carbon-oxygen stratification inside the core of the newborn white dwarf.
Recent  experimental  data  and  updated  values  of  the  astrophysical
$S$-factor  (Arnould et al.  1999) suggest a  substantially  larger rate
than that  derived by Caughlan \& Fowler  (1988), and much closer to the
previously  value  obtained  by  Caughlan  et al.  (1985).  This  yields
larger central  amounts of oxygen as obtained by Salaris et al.  (1997).
However,  as  already  noted by these  authors,  the  internal  chemical
profile  of the white  dwarf  core  depends  not only on the rate of the
$^{12}$C$(\alpha,\gamma)^{16}$O   reaction  but  on  the   treatment  of
convection  as well.  For  instance,  Woosley,  Timmes \& Weaver  (1993)
studied the role of this reaction rate in producing the solar  abundance
set from  stellar  nucleosynthesis  and  concluded  that  the  effective
astrophysical  $S$-factor for the energies  involved  during core helium
burning that best reproduces the observed abundances should be closer to
the rate given by Caughlan \& Fowler (1988).  In their models the Ledoux
criterion  plus an amount of  convective  overshooting  were adopted for
determining the extension of the convective regions.  On the other hand,
the chemical profiles  obtained by Salaris et al.  (1997)  correspond to
an enhanced value for the $^{12}$C$(\alpha,\gamma)^{16}$O  reaction rate
(Caughlan  \&  Fowler  1985)  and  the  adoption  of  the  Schwarzschild
criterion without any overshooting for the convective instability, which
gives an  excellent  agreement  with  the  abundances  of  $^{12}$C  and
$^{16}$O found in the ejecta  SN1987A  (Thielemann,  Nomoto \& Hashimoto
1996).  Thus,   since  we   cannot   disentangle   both   effects,   the
astrophysical  constraints  are  only  set on an {\sl  effective}  cross
section for the $^{12}$C$(\alpha,\gamma)^{16}$O reaction, given the lack
of a reliable  theory of  convection.  Here, in  assuming  the  chemical
profile  computed by Salaris et al.  (1997) we have  assumed  for such a
critical reaction the rate given by Caughlan \& Fowler (1985).  However,
note that a different  reaction rate would obviously produce a different
carbon-oxygen  stratification.  It is well  known  that  changes  in the
internal  composition  of the white dwarf  affect the cooling  timescale
(and thus the rate of change of the period)  because it changes the heat
capacity per mass unit.  As most of the heat capacity of the star is due
to the  contributions  of the  non-degenerate  ions, which is  inversely
proportional to the average atomic number, an enhanced oxygen abundance,
like that found by Salaris et al.  (1997),  implies  less heat  capacity
which, in turn, forces a faster cooling and a larger  $\dot{P}$.  Again,
in terms of Mestel (1952) cooling law:

\begin{equation}
L\simeq - \frac{dU_{\rm ion}}{dt} = - C_{\rm V} M_\star \frac{dT}{dt} 
          \propto \frac{1}{A}
\end{equation} 

\noindent  where $C_{\rm V}$ is the specific  heat and $A$ is the atomic
mass number.  Clearly,  $\tau_{\rm  cool} \propto 1/A$, and consequently
the cooling  proceeds  faster when the oxygen  abundance  is larger and,
according  to  equationn  5, the  secular  rate of change of the  period
becomes  larger.  Because  of this  reason it is  interesting  to have a
quantitative    idea   of   how    such    an    uncertainty    in   the
$^{12}$C$(\alpha,\gamma)^{16}$O  nuclear  reaction rate would affect our
results.  For such a  purpose,  we have  considered  different  internal
carbon-oxygen  stratifications  keeping  the same shape of the  fiducial
profiles but with different central abundances for the carbon abundance,
$X_{\rm C}$, ranging  from  $X_{\rm  C}=0.0$, to $X_{\rm  C}=1.0$.  Such
profiles  are  shown  as  dotted  lines  in  Fig.  \ref{fig:prof-scaled}
together with the fiducial profile, which is shown as a solid line.


\begin{figure} 
\centering
\vspace*{14cm}
\includegraphics{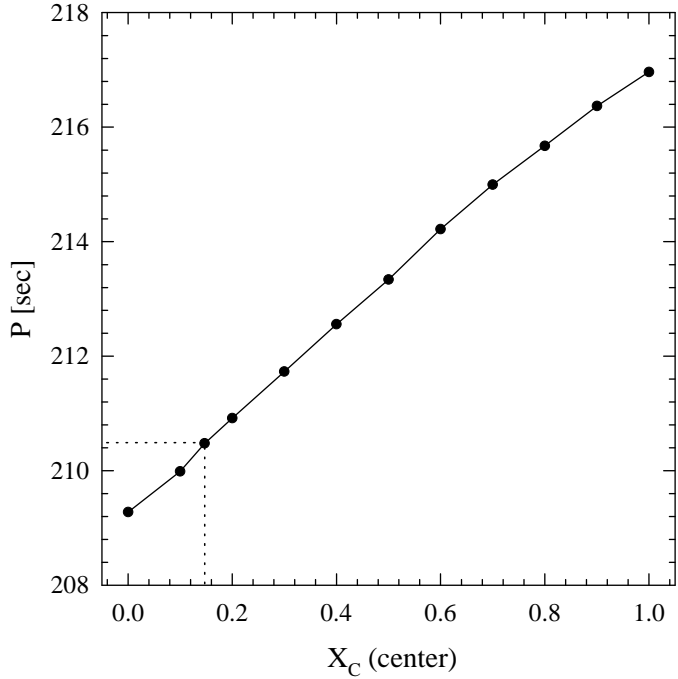}
\caption{The  period of the $l=1$, $k=2$ mode for white dwarf  models of
0.55  $M_{\odot}$  and a $T_{\rm  eff}$ of 11,620 K as a function of the
central carbon  abundance.  The filled dots  represent  computed  models
whereas the short dashed lines  indicate  the  abundance  and the period
corresponding to the fiducial model.}
\label{fig:p-scaled} 
\end{figure}


\begin{figure} 
\centering
\vspace*{14cm}
\includegraphics{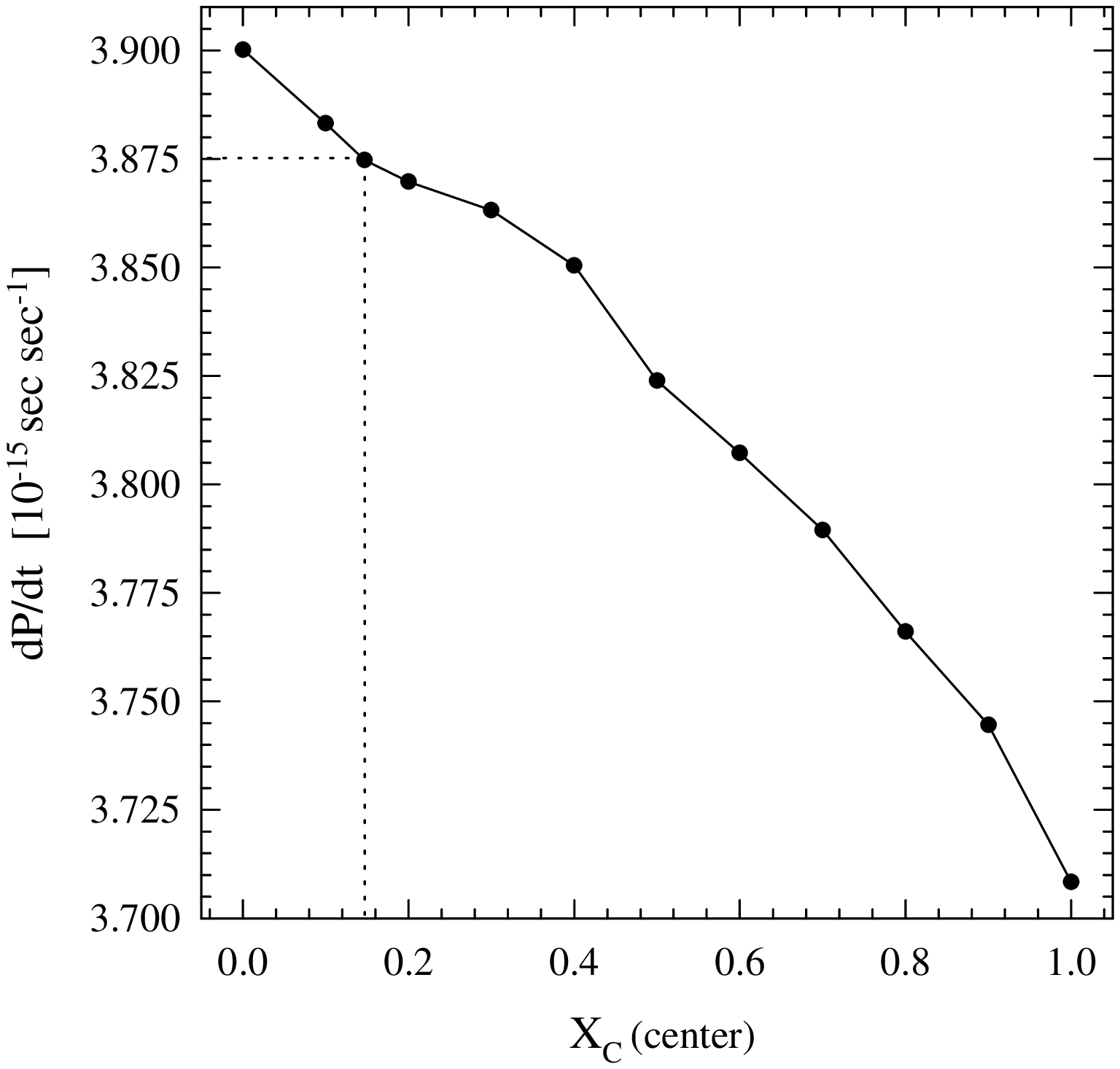}
\caption{The  temporal derivative of the period of the $l=1$, $k=2$ mode
for white dwarf models of 0.55  $M_{\odot}$  and $T_{\rm eff}= 11,620$ K
as a  function  of their  central  carbon  abundance.  The  filled  dots
represent  computed  models whereas the short dashed lines  indicate the
abundance and the temporal derivative of the period corresponding to the
fiducial model.}
\label{fig:pdot-scaled} 
\end{figure}

In Fig.  \ref{fig:p-scaled}  we show the period of the $l=1$, $k=2$ mode
for a white  dwarf  model of 0.55  $M_{\odot}$  and an $T_{\rm  eff}$ of
11,620 K as a function of its  central  carbon  abundance.  Notice  that
changing the carbon  profile in the stellar  core  induces a very modest
change of  $\approx  4\%$ in the  period of the  fiducial  mode.  Such a
change in the period is due to the  following  two  reasons.  First, the
lower the carbon abundance is, the lower the Coulomb interactions in the
stellar  plasma  are.  This, in turn,  produces  slight  changes  in the
structure  of the stellar  core.  Second,  and most  important,  smaller
(larger) carbon central abundances produce steeper (shallower) slopes at
the  outer  boundary  of the  degenerate  core  just at the  base of the
transition  zone,  leading to  different  radial  configurations  of the
modes.  This, in turn, is  important  because in our models, the 215.2 s
mode has  large  amplitudes  in such an  interphase.  Consequently,  any
change in the structure of the  interphase  will affect the structure of
the mode and thus  its  period.  In Fig.  \ref{fig:pdot-scaled}  we show
the  temporal  derivative  corresponding  to the  modes of the  previous
figure.  Here, as it is expected, the lower the central carbon abundance
is, the higher the computed  $\dot{P}$  is.  However,  even  considering
extreme  cases, the  variation of $\dot{P}$ is very  modest:  $\approx 7
\%$.  Thus our results are not severely  influenced by our choice of the
$^{12}$C$(\alpha,\gamma)^{16}$O nuclear reaction rate.

After examining all the possible uncertainties a few words are necessary
to justify the  discrepancy  between the measured  rate of change of the
period of the 215.2 s mode, $\dot{P}= (2.3 \pm 1.4) \times 10^{-15} {\rm
s\, s}^{-1}$, and its computed value for the fiducial  model,  $\dot{P}=
3.9 \times  10^{-15} {\rm s\, s}^{-1}$.  In fact, our computed value for
the fiducial  model,  which is in good  agreement  with the  independent
calculations of Bradley (1998), is slightly beyond the formal  $1\sigma$
error bars.  Kepler et al.  (1995) carefully  discussed the significance
and the  error  bars of all the  previous  determinations  of the  upper
limits to  $\dot{P}$  and argued that there could be some  observational
artifacts due to the  modulation  by the nearby  frequencies.  Moreover,
Costa et al.  (1999) have  recently  shown that a realistic  estimate of
the observational  uncertainties must necessarily include the effects of
all the  periodicities  that  G117-B15A  shows.  However  Kepler  et al.
(2000) already took into account all the  periodicities  and, thus, both
the value of $\dot{P}$ and its reported  error bars can be considered as
safe.  But, on the  other  hand, as  discussed  by  Bradley  (1998)  and
confirmed  by  the  previous  study  of  the  theoretical  uncertainties
performed in this work, a spread of about $\pm 1\times 10^{-15} {\rm s\,
s}^{-1}$   could  be  expected  from  either  the  mode   identification
procedure, the precise observational characteristics (like the effective
temperature  and mass) of G177-B15A and our incomplete  knowledge of the
physical  inputs  involved  in the  calculation  of  $\dot{P}$.  Table 1
summarizes the approximate  theoretical  error budget  obtained from all
the previous  calculations, taking into account realistic  variations of
all the  parameters.  Finally,  it is  worth  mentioning  as  well  that
another source of uncertainty is the  contribution  of the proper motion
of the  G117-B15A  (Pajdosz  1995).  With the current  estimates  of the
proper motion and parallax,  Kepler et al.  (2000) have  concluded  that
the  maximum  contribution  of the  proper  motion  of the  star  to the
observed  rate of change of the period is  $\dot{P}=(9.2\pm  0.5) \times
10^{-16} {\rm s\, s}^{-1}$.  We thus  conclude  that taking into account
the  theoretical  uncertainties,  our  preferred  model  could be safely
considered  as  satisfactory  and that our value for  $\dot{P}$ is fully
consistent with the observed rate of change of the period.

\begin{table}
\centering
\caption{Error budget for the period (in seconds) and for the secular rate of
change of the period, in units of $10^{-15}$s~s$^{-1}$.}
\begin{tabular}{lcc}
\hline
Source & $\Delta P$ & $\Delta\dot{P}$\\
\hline
Mode identification              &  5   & 1.0 \\
$M_\star$                        &  6   & 1.0 \\
$^{12}$C$(\alpha,\gamma)^{16}$O  &  4   & 0.1 \\
$T_{\rm eff}$                    &  2   & 0.2 \\
\hline
\end{tabular}
\end{table}


\section{The   effects   of   an   enhanced   cooling   due   to   axion
emission}\label{sec_axion}


\begin{figure}  
\centering
\vspace*{14cm}
\includegraphics{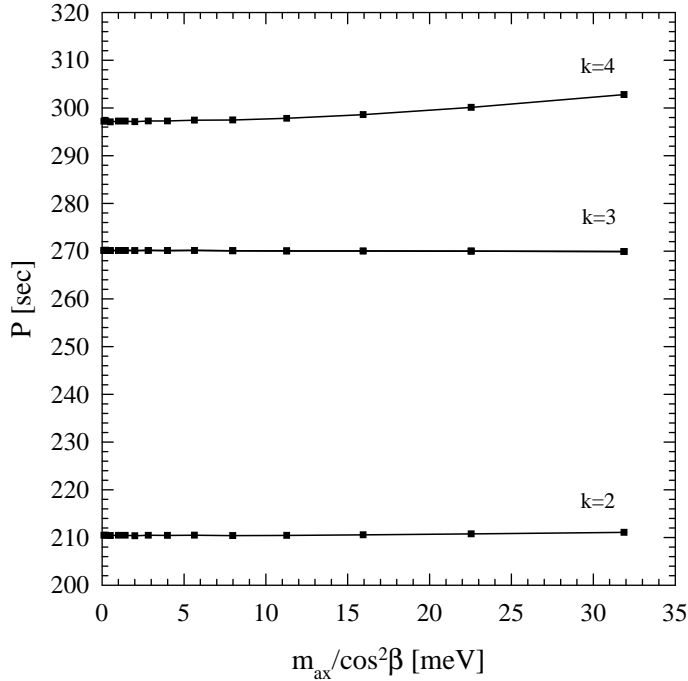}
\caption{The  periods of the $l=1$, $k=2$, 3, and 4 modes  corresponding
to the  fiducial  model at a $T_{\rm  eff}$ of 11,620 K as a function of
the  axion  mass.  Notice  that, in  spite  of the  acceleration  of the
cooling process induced by axion  emission, the period of these modes do
not change significantly.}
\label{fig:p-axion} 
\end{figure}


\begin{figure} 
\centering
\vspace*{14cm}
\includegraphics{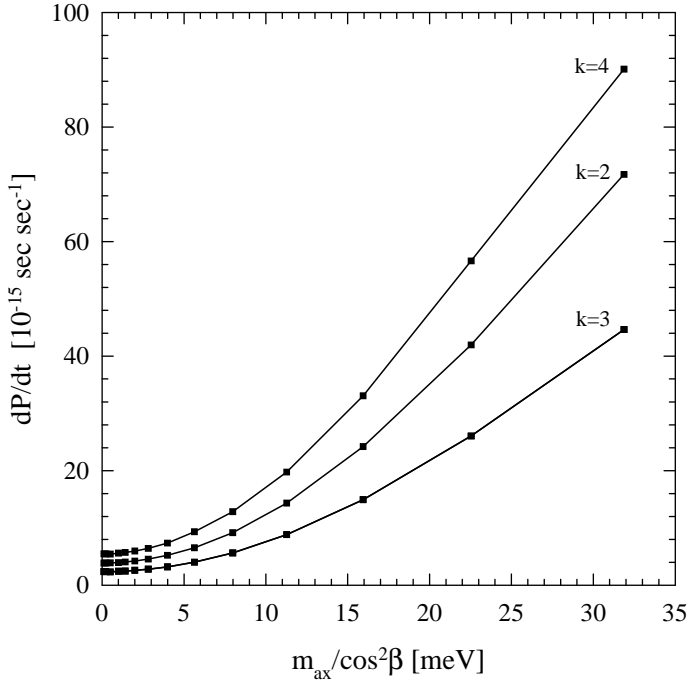}
\caption{The  derivative  of the  periods of the $l=1$,  $k=2$, 3, and 4
modes of the fiducial  model at $T_{\rm  eff}=11,620$  K.  Note that for
high values of the axion mass, as cooling is  strongly  accelerated  the
value of $\dot{P}$ shows a steep increase.}
\label{fig:3pdot-axion} 
\end{figure}


\begin{figure} 
\centering
\vspace*{14cm}
\includegraphics{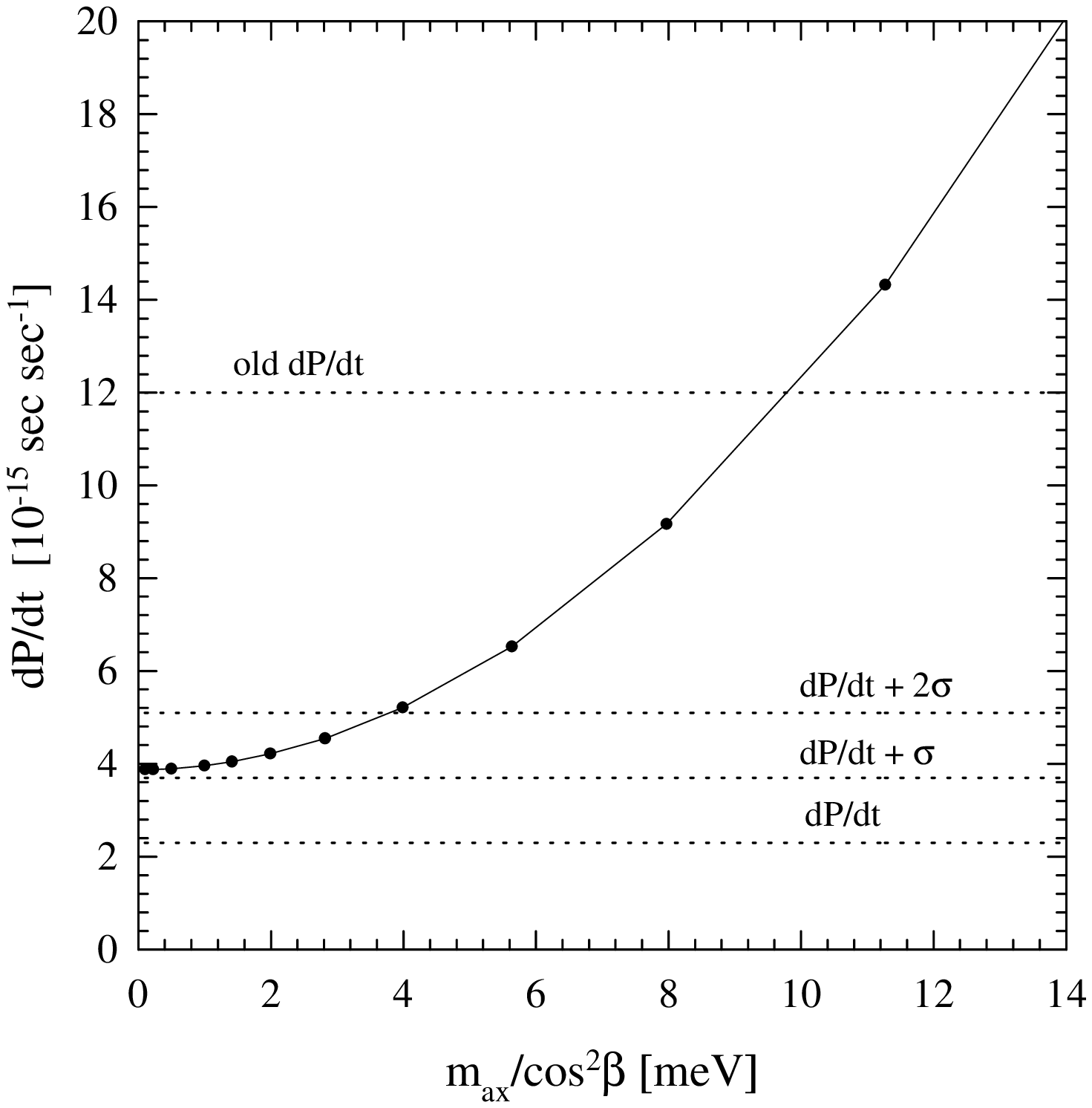}
\caption{The  temporal derivative of the period of the $l=1$, $k=2$ mode
of the fiducial model at $T_{\rm  eff}=11,620$ K.  The horizontal  short
dashed lines indicate the observed value of $dP/dt$, $dP/dt+\sigma$, and
$dP/dt+2\sigma$ from Kepler et al.  (2000).  Also, we show the old value
of $dP/dt$ value  derived by Kepler et al .  (1991).  If we consider two
standard  deviations from the observational  value, we conclude that the
observations   are   compatible   with  an   axion   mass   lower   than
$3.97\cos^{2}{\beta}\;  {\rm  meV}$.  This  is the  main  result  of the
present work.}
\label{fig:pdot-axion} 
\end{figure}

Now, we turn our  attention to compute the effects of axion  emission on
the  evolutionary  timescale of G117-B15A and its effect on the expected
value of $\dot{P}$  for the $l=1$, $k=2$ mode.  In order to do this in a
self-consistent  way we have run an additional set of cooling  sequences
with   different   axion   masses  (and   considered   the   pulsational
characteristics  in the relevant  effective  temperature  range) for our
fiducial model, starting from the same initial  conditions  used for the
models computed  without axion  emission.  This procedure  allowed us to
take into  account  the fact that the axion  emissivity  influences  the
thermal  structure of the white dwarf.  This is  particularly  important
since the cumulative  effect of axions must have reached an  equilibrium
value  before the model  enters into the  instability  strip in order to
avoid an artificial  response  that would not be  representative  of the
real pulsational  characteristics of G117-B15A.  In all the evolutionary
calculations including the axion contribution, we have verified that the
transitory  due to the  inclusion  of axions in the model  occurred  far
before  the star  cools  down to  effective  temperatures  near  that of
G117-B15A.  We have found  that, even  considering  a wide range for the
mass of the axion, the period of the $l=1$, $k=2$, 3, and 4 modes show a
very  small  variation  for the  whole  considered  interval  (see  Fig.
\ref{fig:p-axion},  where we show the  values  of the  period  for these
modes as a function of the assumed mass of the axion).  This is indeed a
very  fortunate  situation  that  allowed us to employ the  procedure of
identifying   first  the   structure  of  the  fiducial   model  without
considering axion emission and then to incorporate the axion emissivity.
This would have not been the case should we have had to identify a white
dwarf  structure  for each  value of the axion  mass.  In such a case we
would  have  had no  fiducial  model,  thus  complicating  our  analysis
enormously.

The  fact  that  the  periods  of  oscillation  at  a  fixed   effective
temperature  are  almost  independent  of the value of axion  mass has a
simple  physical  explanation.  Axion emission  affects the  temperature
profile in the innermost, strongly  degenerate parts of the star.  Thus,
the  structure of the star is only very  slightly  affected by the axion
emissivity.  On its hand, the outer layers are partially  degenerate and
its  structure is dependent  on the  temperature  profile.  But, for the
case we are interested  in, such a profile is largely  determined by the
assumption   of  a  fixed   value   for   the   effective   temperature.
Consequently,  as the modes are dependent upon the mechanical  structure
of the star, the periods are almost unchanged.

In sharp contrast with the small variation of the computed values of the
periods of all the modes with the axion emissivity found previously, the
value of $\dot{P}$ for the three identified modes is extremely sensitive
as it  is  shown  in  Fig.  \ref{fig:3pdot-axion},  where  we  show  the
temporal  derivative of the period for the same range of the mass of the
axion.  As an example,  for the 215.2 s mode, and for the range of axion
masses  considered  here ($0 \leq m_{\rm  ax}/\cos^{2}{\beta}  \leq 32\;
{\rm meV}$), $\dot{P}$ increases by a factor of about 18.  Note that the
$\dot{P}$ value corresponding to the mode with $k=3$ is the smallest one
mode in the whole  considered  axion mass  interval.  This is due to the
fact that this mode is trapped in the outer  hydrogen  envelope  --- see
Bradley, Winget \& Wood (1992) for a physical justification.

Finally, in Fig.  \ref{fig:pdot-axion} we show the value of $\dot{P}$ of
the $l=1$, $k=2$ mode of the fiducial  model at $T_{\rm eff}=  11,620$ K
as a  function  of the mass of the  axion.  The  observational  value of
Kepler et al.  (2000)  is also  shown as a dashed  line.  Also  shown as
dashed lines are the observational errors at $1\sigma$ and $2\sigma$ and
the old value of  $\dot{P}$  (Kepler et al.  1991).  Now we can look for
an upper limit to the axion mass by imposing that the value of $\dot{P}$
should be lower  than the  observed  value  plus two times the  standard
deviation,  that is, lower than $5.1 \times  10^{-15} {\rm s\, s}^{-1}$.
Thus, this  procedure  sets upper limits on the mass of the axion at the
95\%   confidence   level.   From   a   close   inspection   of   figure
\ref{fig:pdot-axion} it is clear that for this to be the case, the axion
mass must be lower than $3.97\cos^{2}{\beta}\;  {\rm meV}$.  This is the
main result of the present  work.  Also,  notice that if we consider the
old $\dot{P}$ value of Kepler et al.  (1991), the upper limit would have
been $\approx  10\cos^{2}{\beta}\; {\rm meV}$ in good agreement with the
results of Isern et al.  (1992).

These results can be easily  explained in terms of a very simple  model.
Isern et al.  (1992)  found  that the  observed  rate of  change  of the
pulsational  period of G117-B15A  $(\dot  P_{\rm  obs})$ and the rate of
change of the period given by the models $(\dot P_{\rm mod})$ when axion
emission is considered are related through the following expression:

\begin{equation}
\frac{L_{\rm phot}+L_{\rm ax}}{L_{\rm phot}}=
\frac{\dot P_{\rm obs}}{\dot  P_{\rm mod}}
\end{equation}

Since we are considering upper bounds at the 2$\sigma$ level in our case
$\dot P_{\rm  obs}=5.1\times  10^{-15} {\rm  s\,s}^{-1}$,  whereas $\dot
P_{\rm  mod}=3.9   \times   10^{-15}  {\rm   s\,s}^{-1}$.  The  observed
luminosity of G117-B15A is $\log(L_{\rm  phot}/L_\odot)=-2.8$ (McCook \&
Sion,  1999).  Thus $\log  (L_{\rm  ax}/L_{\odot})\le  -3.3$.  Since the
axion  emissivity at the relevant  densities and  temperatures  of white
dwarfs  is  dominated  by the  bremsstrahlung  process,  and  since  the
degenerate core of white dwarfs, where most of bremsstrahlung  processes
occur, is essentially isothermal, to a very good approximation (Altherr,
Petitgirard \& del R\'\i o Gaztelurrutia,  1994) the axion luminosity is
given by

\begin{equation}   
L_{\rm ax}/L_\odot\simeq 2.0 \times 10^{22} g_{\rm ae}^2
(M/M_\odot) (T/10^7 {\rm K})^4 
\end{equation}

\noindent  where $T$ is the  temperature  of isothermal  core, which for
G117-B15A is typically  $T\simeq1.2\times  10^7$~K, which yields $g_{\rm
ae}  \le  1.3  \times  10^{-14}$  or,  equivalently,  $\simeq  5\cos^{2}
{\beta}\; {\rm meV}$ at the 95\% confidence level, which is in very good
agreement with the detailed evolutionary calculations reported above.


\section{Summary and conclusions} \label{sec:conclu}

In this paper we have presented a comprehensive study of the pulsational
characteristics  of the  variable DA white dwarf  G117-B15A.  This study
encompasses  the  mode  identification  of  the  observed  periods,  the
characterization  of the  structure,  thickness and  composition  of the
uppermost  regions of the star, the  self-consistent  calculation of the
evolutionary  timescale and,  consequently, of the rate of change of the
215.2  s  period  and,  ultimately,  the  assessment  of the  associated
uncertainties.  All these  required the  construction  of a  pulsational
code  coupled to our  previously  existing  evolutionary  code.  In this
regard our results are in close  agreement with the results  obtained by
other authors (Clemens 1994, Bradley 1998).

Additionally  we have used this ZZ-Ceti star to put  constraints  on the
mass of the axion.  Since G117-B15A is the most stable optical clock yet
known, with a rate of the period of the 215.2 s mode of  $\dot{P}=  (2.3
\pm 1.4) \times  10^{-15}  {\rm s\, s}^{-1}$  (Kepler et al.  2000), the
cooling  timescale  of this white dwarf is well  constrained.  This fact
has allowed us to set up tight  constraints  on any  additional  cooling
mechanism different to the standard ones.  Our results very much improve
those  previously  obtained by Isern et al.  (1992).  In  particular  we
have  obtained an upper bound to the mass of the axion  which is $\simeq
4\cos^{2}{\beta}\;  {\rm meV}$ at the 95\% confidence level.  This upper
limit is a factor of 2.5 smaller than the previously existing limits.

However, from the analysis  performed in the previous  sections it seems
clear that in order to have more  stringent  upper limits to the mass of
axions, we should have a smaller  uncertainty  in the observed  value of
$\dot{P}$,  since the  uncertainties  in the models of white dwarf stars
are clearly of lower  relevance  in this  context.  To this regard it is
important to realize that at the  1$\sigma$  level the  stability of the
dominant period of G117-B15A seems to rule out the existence of the DFSZ
axion, provided that our current knowledge of the origin,  structure and
evolution  of white dwarf stars turns out to be correct.  Thus,  clearly
more observations are required but these  observations are on their way.
Since  the  observational  situation  will  problably  improve  with the
continued  effort of the Whole Earth Telescope  (WET) the  observational
errors will decrease and, thus, there will be the  possibility to set up
even more tight  constraints to the mass of the axion.  Moreover,  there
are other pulsating DA white dwarfs for which an estimate of the rate of
change of the period can be measured like L19-2 and R~548 (Isern et al.,
1993).  Although the quality of the  observations  is not as exceptional
as those of G117-B15A it seems clear that this handful of objects  could
provide  very  valuable  information  and, thus, we  strongly  recommend
continuous coverage in forthcoming WET campaigns.  Finally, there are as
well other --- see the extensive  material  contained in the proceedings
of the WET workshops (Meistas \& Solheim 1996; Meistas \& Moskalik 1998)
--- very well studied DA pulsators, being G29-38 (Kleinman et al., 1998)
perhaps the best  example  among other  ZZ-Ceti  stars, for which a firm
estimate  of their  rate of change of the  period is not yet  available.
Thus, there are more targets to be studied which,  undoubtely, will help
in improving both the theoretical and the  observational  situation.  In
any case G117-B15A seems to be a very promising stellar object to set up
constraints on fundamental physics.


\vskip 1cm
\noindent  
{\bf  Acknowledgments.}  We  would  like  to  warmly  acknowledge  the
comments  by our  anonymous referee  that  allowed us  to improve  the
original version  of this paper.  Part  of this work  was supported by
the   Spanish  DGES   projects  PB98--1183--C03--02,   ESP98-1348  and
AYA2000-1785, by the CIRIT and  by Sun MicroSystems under the Academic
Equipment Grant AEG--7824--990325--SP.



\end{document}